\documentclass[preprint]{aastex}
\begin{document} 
\title{Redshift Evolution of the Merger Fraction of Galaxies in 
CDM Cosmologies}
\author{S. Khochfar and A. Burkert}
\affil{Max-Planck-Institut f\"ur Astronomie, K\"onigstuhl 17, \\
 D-69117 Heidelberg, \\
 Germany}
\email{khochfar@mpia-hd.mpg.de}
\email{burkert@mpia-hd.mpg.de}

\begin{abstract}
We use semi-analytical modelling of galaxy formation to study 
the redshift evolution of the galaxy 
merger fractions and   merger rates in $\Lambda$CDM  and 
quintessence (QCDM) cosmologies, their dependence on 
physical parameters such as the environment, the merger timescale, 
the way major mergers are defined, and the minimum mass of objects 
taken into account.
We find that for a given final halo mass the redshift dependence of
the merger fraction $F_{mg}$ and the resulting merger rate 
can be fitted well by a power law for redshifts $z \lesssim 1$. 
 The normalization $F_{mg}(0)$ and the slope $m$ depend on the final halo
mass. For a given merger timescale $t_{merg}$ and an  assumed maximum
mass ratio $R_{major}$ for major mergers, $F_{mg}(0)$ and $m$ depend 
exponentially on each other. The slope $m$  depends logarithmically on the
ratio of the final halo mass and the minimum halo mass taken into account.
In addition, the local normalization $F_{mg}(0)$ increases 
for larger $R_{major}$ while $m$ decreases. We compare the predicted
merger  fraction
with recent observations  and find that the model cannot reproduce
both the merger index and the normalization at the same time. In
general the model underestimates $F_{mg}(0)$ and $m$ by a factor of 2.
\end{abstract}  
\keywords{ dark matter --- galaxies: formation --- 
galaxies: interactions --- methods: numerical}

\section{INTRODUCTION}\label{int}

Estimating the frequency of mergers in the universe is a challenging task.
Besides the problems of defining  a merger in contrast to 
  an accretion event  and finding such events,  there is also the 
problem of the dependence on the environment,
and the estimate of the merger timescales.  
 For example,  \citet{vd99} find that the merger rate evolves as
$R_{mg}\propto(1+z)^{m}$ with $m=6 \pm 2 $  in rich cluster, whereas the
 merger rate of field galaxies is found to evolve less strongly. 
 In a recent study  \citet{lef00} find $m=3.4 \pm 0.6$ using visually 
classified mergers and  $m=2.7 \pm 0.6$ using close galaxy pairs in 
a population of field galaxies. 
Previous studies found  $m=3.4 \pm 1$ (Carlberg, Pritchet, \& Infante
1994), $m=4 \pm 1.5$
\citep{yee95}, $m=2.8 \pm 0.9$ \citep{pat97}, $m=2.01 \pm 0.52$
\citep{roch99}
 and $m=2.1 \pm 0.5$ \citep{con00}. This spread in the 
values of the merger index $m$ is partly due to different methods used
in deducing  the merger rates \citep[see e.g.][]{pat97,abrah99}. 
Correcting for selection effects in  close pair studies,
\citet{pat00} estimate that approximately $1.1 \% $ of all nearby
galaxies with $-21 \leq M_{B} \leq -18$ are undergoing mergers.

On the theoretical side, Gottl\"ober, Klypin, \& Kravtsov (2001) 
used N-body simulations 
and merger trees based on the Press-Schechter formalism, to derive the
 merger rate. They found $m=3$ for dark matter halos.
In earlier studies of merger rates in N-body simulations \citet{kol99}
found $m=3$ and \citet{gov99} found $m=3.1 \pm 0.2 $ in a
critical universe and $m=2.5 \pm 0.4 $ in an open universe.
%Both studies calculated the merger rate of dark matter halos and did
%not use  semi-analytic methods to estimate the galaxy merger rate. 
 In a previous semi-analytical approach \citet{lc93} calculated the
accretion rate of baryonic cores. They assumed that
each halo has only one baryonic core, neglecting the effect of
multiple cores in a halo.

The purpose of this letter is a detailed investigation of the galaxy
merger fraction and rate.
In the following section the redshift dependence of the merger
fraction and its dependence on the cosmological models, on the
environment represented by the final dark halo mass, on the merger 
timescale, on the minimum mass of observed objects that would be 
identified as  merger components, and on the definition of  major 
mergers are investigated. Besides  allowing a better understanding of
how the merger rates of different observed samples are related, these 
estimates will test cosmological models.     

\section{THE MODEL}\label{model}

We study spatially flat CDM cosmologies with $\Omega_{m}=0.3$,
$\Omega_{\Lambda}=0.7$ and  $H_{0}=65$ km s$^{-1}$Mpc$^{-1}$. 
We also study a quintessence model with  $\Omega_{Q}=0.7$ and an
equation-of-state $w=p/\rho_{Q}=-2/3$ (e.g. Caldwell, Dave, \&
    Steinhardt 1998).
Merger trees of  dark matter
halos were generated using the method described by \citet{somer99},
 which is based on the
 extended Press-Schechter formalism \citep{bond91,bow91}. This method
was modified as proposed by \citet{got01}.  The power spectrum
is obtained from the fitting formula of \citet{bbks86} 
and normalized by $\sigma_{8}$. We use the expressions derived by
\citet{wang98} for the  value of $\sigma_{8}$ and the linear growth
factor.   
The history of a dark matter halo is followed back in time  until the
masses of all its progenitors  fall below a
minimum mass of $M_{min}=10^{10} M_{\odot}$. A progenitor with mass
below $M_{min}$ is assumed  to  inhabit a small galaxy which has
$1/10$ the mass of the  surrounding  dark matter halo. 
Whenever two halos merge the galaxies inside of them merge on a
dynamical friction timescale as described by \citet{kcdw99}.

\section{MERGER FRACTIONS AND RATES}\label{rat}
 From the observational point of view one can 
 either estimate the fraction of visually confirmed mergers 
\citep{lef00} or the
fraction of galaxies in close pairs \citep[e.g][and references
therein]{pat00}.
 To deduce the merger fraction it is necessary to 
correct the observed close pair fraction for background/foreground 
contaminations and to estimate how many of these physical close pairs
 are likely to merge \citep[e.g.][]{yee95,lef00}. Usually one refers
to the merger rate. The connection between the merger rate $R_{mg}(z)$
 and the merger fraction is
\begin{eqnarray}\label{mrate}
R_{mg}(z)=\frac{F_{mg}(z)}{t_{merg}},
\end{eqnarray}
 where $F_{mg}(z)$ denotes the fraction of galaxies at redshift $z$ 
in close pairs which will merge on a timescale shorter than  
$t_{merg}$. Since $t_{merg}$
 depends on the separation of pairs, specifying $t_{merg}$ also
 determines the close pairs. In general, observers measure the
 separation between galaxies in pairs and use the dynamical friction
 estimate to deduce a merger timescale. We calculate the merger fraction
by counting the number of galaxies  at each redshift which are experiencing
 a merger on  a timescale less then $t_{merg}$ and normalizing them to 
the total number of galaxies at this redshift. 
 The merger fraction of galaxies at redshifts $z \lesssim
 1$ is usually approximated by a power law   of the form:
\begin{eqnarray}\label{mr}
F_{mg}=F_{mg}(0)(1+z)^{m},
\end{eqnarray}
where $F_{mg}(0)$ is the normalization to the local merger fraction 
\citep[e.g.][]{lef00}.

\begin{figure}
\plotone{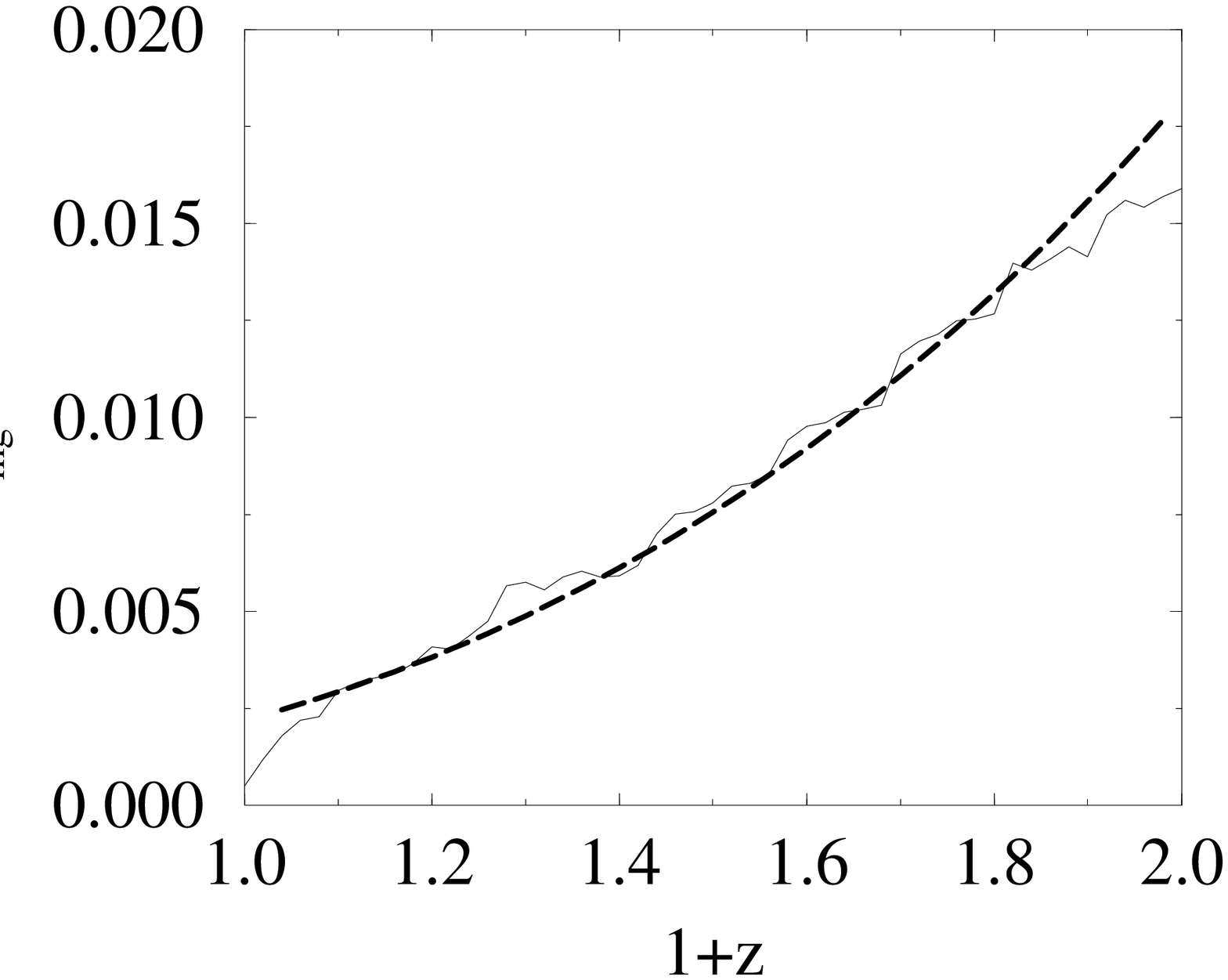}
\caption{ Power law fit to the merger fraction for $M_{0}=5\times 10^{12}
  M_{\odot}$, $R_{major}=4$, $M_{min}=10^{10} M_{\odot}$, and
  $t_{merg}=1$ Gyr in the $\Lambda$CDM model. The solid line
  represents the data and the long dashed line the power law fit 
  for  $z \leq 1$. \label{power} }
\end{figure}

 For our analysis we consider only binary major mergers, which we
 define as mergers with
mass ratio between $R_{major}$ and 1. Fig \ref{power} shows the result
 of a representative  simulation for a halo of mass 
$M_{0}=5\times 10^{12} M_{\odot}$ at
 $z=0$, adopting $M_{min}=10^{10} M_{\odot}$, $R_{major}=4$, and a
 merger timescale of 1 Gyr for the $\Lambda$CDM model. 
We find in all investigated cases that the merger rate and the 
merger fraction as a function 
of redshift can be approximated by a  power law  
at redshifts $z\lesssim 1$, in agreement with the observations. 
At higher redshifts the merger rate flattens, which was also  found by 
 \citet{con00} and \citet{got01}. 

\begin{figure}
\plotone{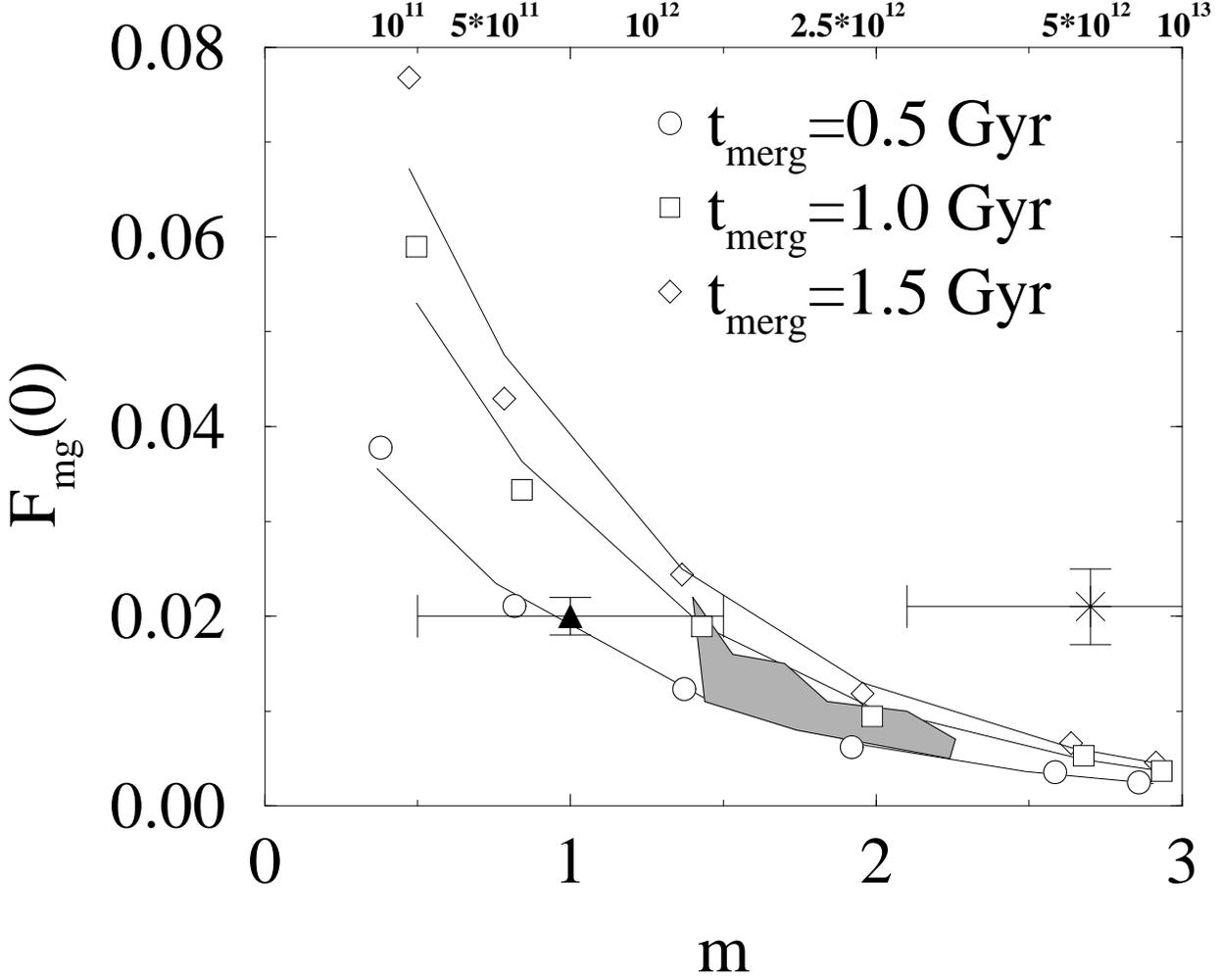}
\caption{Showing the merger fraction $F_{mg}$ at $z=0$ versus the
powerlaw
slope $m$ for major merger events with mass ratios less than
$R_{major}=4$.
The data points correspond to different values of the merger timescale 
$t_{merg}$ and final halo mass $M_{0}$. 
 In the upper part of the figure the final halo masses in units of $M_{\odot}$
are indicated. Halos of the same mass $M_{0}$ have roughly the same value of
$m$. The curves show exponential laws, fitted to the data
for $t_{merg}=0.5$ Gyr, 1 Gyr and 1.5 Gyr 
respectively. The  shaded region represents the  Press-Schechter weighted
average  merger fraction of galaxies in dark halos for  the same
range of $t_{merg}$ as mentioned above.  
 The star indicates $F_{mg}(0)$ and $m$ as  estimated by
\citet{lef00}. The triangle is the result from the combined CFGRS
and CNOC2 data (R. Carlberg, private communication).\label{mass} }
\end{figure}

In  general, a range of final halo masses will contribute to the
merger events seen in observational surveys.
To take this into account and to estimate environmental effects we choose
six different halo masses $M_{0}$  at redshift $z=0$ 
($M_{0}=10^{11}, 5 \times 10^{11}, 10^{12},
 2.5 \times 10^{12}, 5\times 10^{12}, 10^{13}$; in units of $M_{\odot}$). 
Fig. \ref{mass} shows the
dependence of $F_{mg}(0)$ and $m$ on $M_{0}$ and $t_{merg}$. For
increasing $M_{0}$,
$F_{mg}(0)$ decreases and $m$ increases systematically.
This trend is consistent with the findings of \citet{vd99}. 
Varying $t_{merg}$ corresponds to different definitions of close pairs.
The three curves in fig. \ref{mass} are exponential laws of the form
$F_{mg}(0)=c_{1}\exp(c_{2}m)$, fitted to  
  the merger fractions  for different
$t_{merg}$. The parameters used to fit the data points  are
$c_{1}=0.058$ and  $c_{2}=-1.23$, 
  $c_{1}=0.107$ and $c_{2}=-1.34$, and $c_{1}=0.137$ and $c_{2}=-1.42$ for
$t_{merg}$ equal to 0.5 Gyr, 1 Gyr and 1.5 Gyr, respectively.
 In the same environment, that is the same final halo mass $M_{0}$, 
 $F_{mg}(0)$ increases with increasing 
merger timescale as binaries with larger separations are included.
 The merger index $m$ shows only weak variation. 

\begin{figure}
\plotone{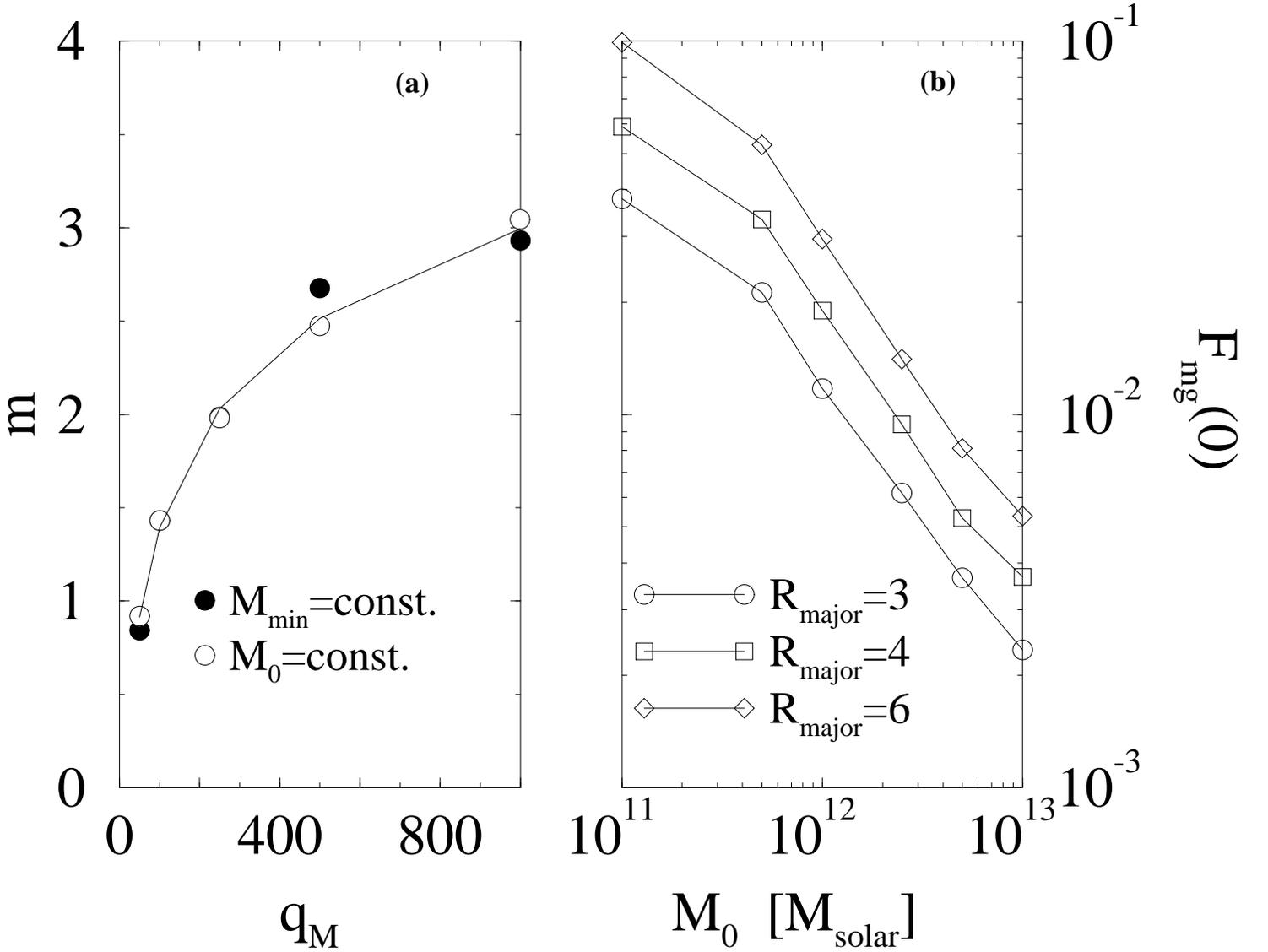}
\caption{Panel (a) shows the dependence of the merger index $m$
on the mass ratio $q_{M}$.
 The points are fitted by
$m=c_4\ln(M_{r})+c_{5}$  with $c_{4}=0.69 \pm 0.09$ and $c_{5}=-1.77
\pm 0.47$. Panel (b) of this figure shows the local merger fraction
for  cases with 
$M_{0}=10^{11} m_{\odot}$, $M_{min}=10^{10} m_{\odot}$, and varying 
$R_{major}$ ($R_{major}=3,4,6$). 
Larger values of $R_{major}$ show larger values of $F_{mg}(0)$.  
 Both results
refer to   $R_{major}=4$.
The graphs in (a) and (b) refer to the  $\Lambda$CDM model and 
$t_{merg}=1$ Gyr. \label{majmer}} 
\end{figure}

For computational reasons mergers are only resolved above a minimum
mass $M_{min}$. Mergers below this mass are neglected. This corresponds 
to observations with a magnitude limited sample of galaxies.
 The graphs in fig. \ref{majmer}a
 show the dependence of the merger index $m$ on $q_{M}=M_{0}/M_{min}$. 
The filled circles are the results of merger trees with constant 
$M_{min}=10^{10} M_{\odot}$ and varying  $M_{0}$. We compare these
 results with simulations for constant
$M_{0}=10^{11} M_{\odot}$ and varying $M_{min}$, represented by open
circles. The value of $m$ depends only on the ratio $q_{M}$ as 
$m=0.69\ln(q_{M})-1.77$.

Another important question is the influence of the
definition of major mergers on the merger rate.
The  graphs in fig. \ref{majmer}b  show  the
dependence of $F_{mg}(0)$ on different values of $R_{major}$.
 An event is called major merger if the mass ratio of the merging  
galaxies is below $R_{major}$ and larger than 1. 
As $R_{major}$ increases, $F_{mg}(0)$ increases.
We also find that the merger index $m$ stays roughly constant for
 low $R_{major}$ and decreases at larger $R_{major}$.
A decrease in $m$ with larger $R_{major}$ has also been
reported by \citet{got01}. It is a result of the adopted minimum mass
for merger events. The detectable 
amount of major mergers with large mass ratios decreases faster with
redshift than for equal mass mergers, since the small masses drop
faster below the minimum mass.
 In observed samples of close pairs \citet{roch99} and \citet{pat00}
found that $F_{mg}(0)$ increased  when they allow for larger
$R_{major}$, which agrees with our predictions.

How do the theoretical models compare to the observations? 
The star in fig. \ref{mass} is the measured merger fraction for field
galaxies
by \citet{lef00}, who used $R_{major}=4$ and the local  merger fraction of
 \citet{pat97}. They identified close pairs as  those which merge on a 
timescale less then $t_{merg}=1$ Gyr. To compare this merger fraction
 with our estimates one needs to take into account that the dark halos
 of field galaxies can vary over a range of masses and that the merger
 timescale is subject to large uncertainties. 
We therfore weighted the different merger fractions of our
 sample of field galaxies with halo masses $M_{0}$
between $5\times10^{11} M_{\odot}$ and  $5 \times 10^{12}
 M_{\odot}$ according to the 
Press-Schechter predictions. The merger index $m$ and the local
 merger fraction 
$F_{mg}(0)$ for different $M_{0}$ were calculated using the  fitting
formulae as shown by  the graphs in fig. \ref{mass} and fig.
\ref{majmer}a.
We varied the the range of halo masses contributing to the sample by
changing the lower bound of halo masses  from $5\times10^{11}
 M_{\odot}$ to $2.5\times10^{12} M_{\odot}$ and changed $t_{merg}$
within the range of 0.5 - 1.5 Gyr. The results of this reasonable
parameter range lie inside the shaded region in fig. \ref{mass}. Results
 for larger $t_{merg}$ correspond to  the upper part of the region and
 those for  larger  halo masses lie in the right part of the region. 
A comparison of our results with the observations shows, that the
predicted merger  index $m$ and the normalization $F_{mg}(0)$ are a 
factor 2 smaller than observed. 

As a possible solution to this problem we have studied a quintessence
model with $w=-2/3$. The  QCDM model shows a  shallower increase in the
comoving number density  of mergers with redshift than the
 $\Lambda$CDM model. There is however not a significant difference in 
the merger fractions. This results from the fact that the difference
in the comoving number density is
 compensated by the length of the redshift range contributing to
 the merger fractions. The QCDM universe with an age of $\sim
1.36\times 10^{10}$ years is younger than the adopted $\Lambda$CDM
universe with an age of $\sim 1.45\times10^{10}$ years, which is the
reason why the same $t_{merg}$ refers to a larger redshift range in
the QCDM universe.
 This result also emphasizes, that it will not be possible to
break  the degeneracy of these models by measuring merger rates.
Comparing \citet{lef00} results with those obtained from
the combined Caltech Faint Galaxy Redshift Survey (CFGRS) and Canadian
Network for Observational Cosmology field galaxy survey (CNOC2)
 (R. Carlberg, private communication), which includes also minor
majors, reveals that
including minor mergers leads to a smaller merger index $m$ which is
 consistent with the predictions of our simulations. It is therefore not
surprising that these two observed merger indices differ. 

\section{DISCUSSION AND CONCLUSIONS}

Using semi-analytical modelling we recover a power law for the
evolution of the merger rates and fractions at $z\leq 1$, as has 
been reported in earlier work.
 Varying the final mass $M_{0}$, the local
merger fraction $F_{mg}(0)$  shows an
exponential dependence on the merger index $m$ of the form
$F_{mg}=c_{1}\exp(c_{2}m)$. The actual values of  the
parameters $c_{i}$ depend mainly on the merging timescale and 
on the definition of major mergers.
 Our predictions that $m$ will increase and $F_{mg}(0)$ will decreases in
more massive environments is in qualitative agreement with
observations. The merger index $m$ depends on the environment through the
mass ratio $q_{M}$. The logarithmic function $m=c_{4}\ln(q_{M})+c_{5}$
fits the data well.
 We find a similiar behavior as \citet{pat00}, which have shown that if
they extend their galaxy sample to fainter magnitudes the local merger
fraction rises. In addition, we also find that the merger index
decreases with higher mass ratios. This is also beeing found by
comparing the results of the combined CFGRS and CNOC2 sample with
those of \citet{lef00}. The adopted QCDM model does not show any significant
difference to the $\Lambda$CDM model. Therefore it is not possible to 
distinguish between these two models by  measuring the merger rate of 
galaxies.  
 
Our model predicts values for $F_{mg}(0)$ and $m$ which are too small
 by a factor of 2 compared with the predictions by \citet{lef00} who used
 the local merger fraction estimate of
\citet{pat97} which was derived with a different definition of major
mergers than theirs. As we have shown, the definition of a major merger
is crucial for the expected merger fraction. Our results indicate that the 
local merger fraction $F_{mg}(0)$ for the galaxy sample of
\citet{lef00} who used $R_{major}=4$  must be less 
than the value measured by  \citet{pat97} who used a larger
$R_{major}$.  A smaller value of $F_{mg}(0)$  would lead to an even larger
discrepancy in $m$ compared to our results.
 Another issue might  
 be  observational errors, like projection effects or 
uncertanties in the merger timescale estimates. 
Our results  strongly emphasize that the comparison of merger
 fractions deduced
from different samples and with alternative techniques is questionable
 if the adopted mass range and the definitions of close pairs and 
major mergers are not the same.
\newline

We thank  Ray Carlberg for pointing out the results from the 
CFGRS and CNOC2 data and Stefan Gottl\"ober for usefull comments.

\clearpage

\end{document}